\documentclass[a4paper,fleqn,usenatbib]{mnras}

\usepackage{mathptmx}

\usepackage[T1]{fontenc}
\usepackage{ae,aecompl}

\usepackage{graphicx}	
\usepackage{amsmath}	
\usepackage{amssymb}	
\usepackage{pdflscape}  
\usepackage{bm}

\title[A test of cosmological principle using SNIa]{A tomographic test of cosmological principle using the JLA compilation of type Ia supernovae}

\author[Z. Chang, H.-N. Lin, Y. Sang and S. Wang]{
Zhe Chang$^{1,2}$
Hai-Nan Lin,$^{3}$
Yu Sang,$^{1,2}$\thanks{Corresponding author: sangyu@ihep.ac.cn}
and
Sai Wang$^{4}$\thanks{Corresponding author: wangsai@ihep.ac.cn}
\\
$^{1}$Institute of High Energy Physics, Chinese Academy of Sciences, Beijing 100049, China\\
$^{2}$School of Physical Sciences, University of Chinese Academy of Sciences, Beijing 100049, China\\
$^{3}$Department of Physics, Chongqing University, Chongqing 401331, China\\
$^{4}$Department of Physics, The Chinese University of Hong Kong, Shatin, N.T., Hong Kong 999077, China\\
}

\date{Accepted XXX. Received YYY; in original form ZZZ}

\pubyear{2018}

\begin{document}
\label{firstpage}
\pagerange{\pageref{firstpage}--\pageref{lastpage}}
\maketitle

\begin{abstract}
We test the cosmological principle by fitting a dipolar modulation of distance modulus and searching for an evolution of this modulation with respect to cosmological redshift. Based on a redshift tomographic method, we divide the Joint Light-curve Analysis compilation of supernovae of type Ia into different redshift bins, and employ a Markov-Chain Monte-Carlo method to infer the anisotropic amplitude and direction in each redshift bin. However, we do not find any significant deviations from the cosmological principle, and the anisotropic amplitude is stringently constrained to be less than a few thousandths at $95\%$ confidence level.
\end{abstract}




\section{Introduction}\label{sec:introduction}
The cosmological principle, hypothesizing that the universe is statistically homogeneous and isotropic on large scales, is one of the foundations of modern cosmology \citep{Weinberg1972}. Based on it, a standard model of cosmology, namely the cosmological constant plus cold dark matter ($\Lambda$CDM) model, has been well established in past decades, and all the six base cosmological parameters have been measured to a precision of few percents using cosmological observations, such as the cosmic microwave background (CMB) observed by Wilkinson Microwave Anisotropy Probe (WMAP) \citep{Bennett2013,Hinshaw2013} and Planck \citep{Planck2015XIII,Planck2015XX} satellites. However, there still exist several longstanding challenges to the $\Lambda$CDM model (see reviews in Refs.~\citep{Perivolaropoulos2008,Perivolaropoulos2011,Mariano2013,Bull:2015stt}).

Among these challenges, one is referred to cosmological large-scale ``anomalies'', such as a spatial variation of the electromagnetic fine structure constant \citep{Webb2011,King2012,Mariano2012,Molaro2013}, a large-scale alignment of the quasar polarization vectors \citep{Hutsemekers2001,Hutsemekers2005}, an alignment of CMB low-$\ell$ multipoles \citep{Lineweaver1996,Tegmark2003,Bielewicz2004,Frommert2010,Copi2010}, a parity asymmetry in the angular power spectrum of CMB temperature anisotropies \citep{Kim2010a,Kim2010b,Kim2011,Gruppuso2011,Zhao2014}, and a hemispherical power asymmetry of the CMB temperature anisotropies \citep{Eriksen2004,Hansen2004,Bennett2013,Akrami2014,Planck2013XXIII,Quartin2015}, and so on. Due to explicit dependence on spatial orientations, these anomalies might imply deviations from the homogeneity and isotropy of the universe. Or equivalently, there may be a preferred direction in the universe. To describe the deviations, several theoretical models have been proposed, for example, Bianchi-I cosmology \citep{Campanelli2006,Campanelli2007,Campanelli2011,Schucker2014}, Bianchi-III and Kantowski-Sachs metrics \citep{Koivisto2011}, Randers-Finsler cosmology \citep{Chang2013a,Chang2014b,Chang2014a,LiXin2014a,LiXin2017b,LiXin2017a}, etc.

Besides studying the accelerating expansion of the universe \citep{Riess1998,Perlmutter1999}, the supernovae of type Ia (SNIa)  as standard candles have been widely used to explore the inhomogeneity and anisotropy of the universe \citep{Schwarz2007,Gupta2008,Antoniou2010,Blomqvist2010,Colin2011,Cai2012,Mariano2012,Cai2013,Kalus2013,Zhao2013,Chang2014b,Heneka2014,WangJS2014,
Yang2014,Chang2014a,Chang2015a,Javanmardi2015,LiXin2015a,Lin2016a,Lin2016b,LiXin2017b,LiXin2017c}.
The dark energy fluctuations were found to be vanishing when the Union2 compilation was used to study the angular covariance function of SNIa magnitude residuals \citep{Blomqvist2010}. A ``residual'' statistic was constructed to search for a preferred direction in the Union 2 dataset, and the SNIa with $z<0.5$ were indicated to be inconsistent with the $\Lambda$CDM model at 2--3$\sigma$ CL \citep{Colin2011}. The hemisphere comparison method \citep{Schwarz2007,Antoniou2010,Cai2012,Kalus2013,Yang2014,Chang2014a,Bengaly2015,Chang2015a,Lin2016b} and the dipolar modulation method \citep{Mariano2012,Cai2013,WangJS2014,Salehi2016} were widely used to identify the preferred directions using the Union2 dataset. The power spectrum calculation method was used to study the Union 2.1 dataset, and implied a dipolar anisotropy at less than 2$\sigma$ \citep{Ghodsi2017}. The up-to-date compilation of SNIa is the ``Joint Light-curve Analysis'' (JLA) compilation \citep{Betoule2014}, which improves the accuracy of the photometric calibration, and reduces the uncertainties in the SNIa light-curve model. The Markov-Chain Monte-Carlo (MCMC) method was applied to constrain the anisotropic amplitude and direction in dynamical dark energy models using the JLA sample \citep{Lin2016a}. The anisotropic amplitude has an upper bound of $1.98\times 10^{-3}$ at 95 \% CL for the dipole-modulated $\Lambda$CDM model. A test on dipolar distance duality relation using JLA compilation implied an isotropic universe at 2$\sigma$ CL \citep{LiXin2017c}. More recently, the constraints on the anisotropy of Bianchi-I cosmology, also using the JLA dataset, were found to be consistent with an isotropic universe \citep{WangYY2017}. In JLA data at low redshift range ($z \leq 0.1$), no significant evidence for anisotropy was found \citep{Andrade:2017iam}.

The redshift dependence of the SNIa light-curve parameters has been investigated extensively \citep{Marriner2011,Mohlabeng2014,WangS2013,WangS2014b,Shariff2015,LiM2016,WangS2017}. The stretch-luminosity parameter $\alpha$ is consistent with a constant over the whole redshift range, while the color-luminosity parameter $\beta$ has significant redshift dependence. The redshift evolution of $\beta$ was found in the SDSS-II sample of SNIa \citep{Marriner2011}.  A linear redshift evolution of $\beta$, rather than a constant $\beta$, was found in the Union~2.1 compilation \citep{Mohlabeng2014}. When the statistical and systematic uncertainties were included, a strong
evidence for the deviation from a constant $\beta$ was found using the SNLS3 sample \citep{WangS2013}. Considering the effects of $\beta$ evolution on dark energy models, $\beta$ was showed to deviate from a constant at 6$\sigma$ CL using the SNLS3 sample \citep{WangS2014b}. Two typical parameterizations of $\beta$ were considered in the JLA compilation, and a sharp drop of $\beta$ was showed at the redshift $z=0.662$ at around 4$\sigma$ CL \citep{Shariff2015}. Using a redshift tomography method, a study showed that $\alpha$ is always consistent with a constant, while $\beta$ has a significant trend of decreasing at high redshifts at 3.5$\sigma$ CL in the JLA compilation \citep{LiM2016}. If a flux-averaging technique was applied to the JLA sample, one can significantly reduce the redshift evolution of $\beta$ \citep{WangS2017}.

The redshift evolution of color-luminosity parameter $\beta$ may have a significant impact on testing the homogeneity and isotropy of the universe using SNIa. However, this effect has not yet been took into account in existing studies that test the cosmological principle. In this work, we test the cosmological principle by applying the redshift tomographic method to take the $\beta$ evolution into account. To be specific, we divide the whole JLA compilation into several subsamples with different redshift bins, and then estimate the model parameters including the anisotropic amplitude and direction for each subsample. We wonder if there are evidences for the deviations from the cosmological principle, and their evolution with the redshift.

The rest of the paper is arranged as follows. In section \ref{sec:methodology}, we describe the theoretical model, observational data, and data analysis method. In section \ref{sec:results}, we demonstrate our results of the constraints on deviations from the cosmological principle. The conclusions and discussions are given in section \ref{sec:conclusions}.

\section{Methodology}\label{sec:methodology}
In a spatially flat Friedmann-Robertson-Walker space-time, a luminosity distance to a SNIa is given by
\begin{equation}\label{eq:luminosity-distance}
d_\textrm{L}=\frac{1+z_{\mathrm{hel}}}{H_0}\int_0^{z_{\mathrm{cmb}}}\frac{dz}{E(z)}\ ,
\end{equation}
where $H_0$ is the Hubble constant, $z_{\mathrm{cmb}}$ and $z_{\mathrm{hel}}$ denote the CMB rest-frame redshift and the heliocentric redshift, respectively. \footnote{ Compared to the standard luminosity distance, given by
\begin{equation}\label{eq:standard-luminosity-distance}
d^\textrm{std}_\textrm{L}=\frac{1+z}{H_0}\int_0^{z}\frac{dz}{E(z)}\ ,
\end{equation}
Eq.(\ref{eq:luminosity-distance}) contains two different redshifts, among which $z_{\mathrm{hel}}$ is used in the fits of light curve model (e.g., SiFTO \citep{Conley:2008xx} and SALT2 \citep{Guy:2007dv}) and $z_{\mathrm{cmb}}$ includes the corrections of the nearby SNIa peculiar velocities relative to the Hubble flow \citep{Conley2011}. The difference between $z_{\mathrm{hel}}$ and $z_{\mathrm{cmb}}$ is only important for some of the lowest-$z$ SNIa \citep{Sullivan2011}. Both SNLS and JLA samples use Eq.(\ref{eq:luminosity-distance}) in CosmoMC for cosmological fits. }
In the $\Lambda$CDM model, the function $E(z)$ is given by
\begin{equation}\label{eq:E-z}
E(z)=\left[\Omega_\textrm{m}(1+z)^3+(1-\Omega_\textrm{m})\right]^{1/2}\ ,
\end{equation}
where $\Omega_\textrm{m}$ is the matter density today.
The distance modulus in an isotropic universe is defined as
\begin{equation}\label{eq:distance-modulus}
\bar{\mu}_{\textrm{th}}=5 \textrm{log}_{10}\frac{d_\textrm{L}}{10 \textrm{pc}}\ .
\end{equation}

The anisotropy of the universe, derived from either an anisotropic repulsive force of dark energy \citep{Armendariz-Picon2004,Koivisto2008,Salehi2016} or a preferred-directional background space-time \citep{Chang2013a,Chang2013b,LiXin2013a,LiXin2015a,Schucker2014},  is assumed to induce a dipolar modulation to the  distance modulus, namely
\begin{equation}\label{eq:dipole-distance-modulus}
\mu_{\textrm{th}}=\bar{\mu}_{\textrm{th}}(1+A_\textrm{D}(\bm{\hat{n}\cdot\hat{p}})),
\end{equation}
where $A_\textrm{D}$ is a dipolar amplitude, the unit vector $\bm{\hat{n}}$ describes a dipolar direction in the sky, and the unit vector $\bm{\hat{p}}$ is a line-of-sight vector towards a SNIa. We use the Galactic coordinate to parameterize $\bm{\hat{n}}$ as $\bm{\hat{n}}=\textrm{cos}(b)\textrm{cos}(l)\bm{\hat{i}} +\textrm{cos}(b)\textrm{sin}(l)\bm{\hat{j}} +\textrm{sin}(b)\bm{\hat{k}}$, where $l$ and $b$ are Galactic longitude and latitude, and $\bm{\hat{i}}$, $\bm{\hat{j}}$, $\bm{\hat{k}}$ are three unit vectors in the Cartesian coordinate.
Therefore, the anisotropy of the universe is completely described by a parameter space including three parameters ($A_D$, $b$, $l$).
Correspondingly, the line-of-sight orientation of $i$th supernova is given as $\bm{\hat{p}}_i=\textrm{cos}(b_i)\textrm{cos}(l_i)\bm{\hat{i}} +\textrm{cos}(b_i)\textrm{sin}(l_i)\bm{\hat{j}} +\textrm{sin}(b_i)\bm{\hat{k}}$, in which the Galactic coordinate is ($b_i$, $l_i$).

The empirical observation of SNIa gives the distance estimation using a linear model, under an assumption that all the SNIa have a same intrinsic luminosity for all redshits if they are identical on color, shape and galactic environment \citep{Betoule2014}. This model, used by most similar cosmological analyses, yields the distance modulus as
\begin{equation}\label{eq:jla-distance-modulus}
\mu_{\textrm{obs}}=m_B^*-(M_B-\alpha\times X_1 +\beta\times C )\ .
\end{equation}
Here $m_B^*$ is the observed peak magnitude in rest-frame $B$ band, $X_1$ is the time stretching of the light-curve, and $C$ is the supernova color at maximum brightness. These three quantities can be obtained from a fit to the light-curve model of SNIa. In addition, $\alpha$, $\beta$ and $M_B$ are three nuisance parameters, where the stretch-luminosity parameter $\alpha$ is identical for all the SNIa, the color-luminosity parameter $\beta$ and the absolute magnitude $M_B$ depend on the host galaxy properties \citep{Sullivan2011,Johansson2013,Betoule2014}. In cosmological analysis, $\beta$ is usually taken as a constant for all the SNIa, since it is believed to have no significant effect on cosmological results. In fact, however, $\beta$ is found to have a significant trend of decreasing at high redshift, at around $3.5\sigma$ CL \citep{LiM2016}. In this work, we will study the impacts of $\beta$'s evolution on inference of model parameters. The absolute magnitude $M_B$ is assumed to be related to the host stellar mass $M_\textrm{stellar}$ through $M_B=M_B^1$ if $M_\textrm{stellar}<10^{10} M_\odot$ while $M_B=M_B^1 + \Delta_M$ if otherwise. Here $M_B^1$ and $\Delta_M$ are two nuisance parameters which are marginalized in cosmological fittings.\footnote{The JLA provides two versions of likelihood code. One is written in C++, and in it $M_B^1$ and $\Delta_M$ are free parameters. As a plugin of CosmoMC, the other is written in Fortran, and in it $M_B^1$ and $\Delta_M$ are marginalized using an algorithm described in Appendix C of Ref.~\citep{Conley2011}. We use the latter one in this work.}

To constrain the anisotropy of the universe, the JLA compilation \citep{Betoule2014} is adopted in this work. Covering a redshift range 0.01 $<z<$ 1.3, the JLA compilation assembles the SNIa samples of all three seasons from SDSS-II (0.05 $<z<$ 0.4), of three years from SNLS (0.2 $<z<$ 1), a few very high redshift (0.7 $<z<$ 1.4) samples from HST, and several low redshift ($z<$ 0.1) programs. All of 740 SNIa are spectroscopically confirmed with high-quality light curves.

To study the evolution of model parameters with redshift, we employ a redshift tomographic method to analyze the JLA sample. To be specific, the JLA sample is binned into several subsamples according to cutting redshifts $z_{\mathrm{cut}}$. For each subsample, the model parameters are assumed to be piecewise constants. In realistic operation, we study the following nine ways of tomography, namely
\begin{description}
  \item[1)] 2 bins with $z_\textrm{cut} =$ 0.1
  \item[2)] 2 bins with $z_\textrm{cut} =$ 0.2
  \item[3)] 2 bins with $z_\textrm{cut} =$ 0.3
  \item[4)] 2 bins with $z_\textrm{cut} =$ 0.4
  \item[5)] 2 bins with $z_\textrm{cut} =$ 0.5
  \item[6)] 2 bins with $z_\textrm{cut} =$ 0.6
  \item[7)] 3 bins with $z_\textrm{cut} =$ 0.15, 0.3
  \item[8)] 4 bins with $z_\textrm{cut} =$ 0.15, 0.25, 0.5
  \item[9)] 5 bins with $z_\textrm{cut} =$ 0.1, 0.2, 0.3, 0.6
\end{description}
The number of SNIa in each subsample is denoted by $N_{\mathrm{SN}}$. For the last three cases, $z_\textrm{cut}$ are chosen to ensure roughly equal number of SNIa in each subsample. We do not divide the JLA sample in high-redshift range $(0.6<z<1.3)$ with a $z_\textrm{cut}$, due to a small number of SNIa in this redshift range.

For each subsample, the model parameters can be estimated by minimizing the $\chi^2$ function
\begin{equation}\label{eq:chi-squared}
\chi^2=(\bm{\mu}_\textrm{obs}-\bm{\mu}_\textrm{th})^\dag\textrm{C}^{-1}(\bm{\mu}_\textrm{obs}-\bm{\mu}_\textrm{th}),
\end{equation}
where $\bm{\mu}_\textrm{th}$ is the modulated distance modulus given in Eq.~(\ref{eq:dipole-distance-modulus}), $\bm{\mu}_\textrm{obs}$ is observed distance modulus given in Eq.~(\ref{eq:jla-distance-modulus}), and $\textrm{C}$ is a covariance matrix among SNIa in this subsample.
Here we construct $\textrm{C}$ of each subsample from the full covariance matrix of all 740 SNIa provided in Ref. \citep{Betoule2014}. The covariance matrix is a sum of statistical and systematic uncertainties, including error propagation of light-curve fit uncertainties, calibration uncertainty, light-curve model uncertainty, bias correction uncertainty, mass step uncertainty, and uncertainties in peculiar velocity corrections and contamination of non-Ia supernovae. Note that $\textrm{C}$ depends on nuisance parameters $\alpha$ and $\beta$ and is recomputed at each step when minimizing the $\chi^2$ function.

For each subsample, we use the Markov-Chain Monte-Carlo (MCMC) method to infer the allowed parameter space. To be specific, we employ a modified version of the publicly available CosmoMC package \citep{Lewis2002}, and the likelihood function is given by $\mathcal{L}\propto\exp(-\chi^2/2)$. Following Ref. \citep{Betoule2014}, we set a fixed fiducial value of Hubble constant, namely $H_0=70\textrm{ km s}^{-1} \textrm{Mpc}^{-1}$. The full parameter space consists of the matter density today $\Omega_\textrm{m}$, dipolar amplitude $A_\textrm{D}$, Galactic longitude $l$ and latitude $b$ of the dipole, and two nuisance parameters $\alpha$ and $\beta$. The other two nuisance parameters $M_B^1$ and $\Delta_M$ are marginalized here. Note that the anisotropic amplitude $A_D$ is set to be positive a priori.

\section{Results}\label{sec:results}

Our results are listed in Tab.~1, and depicted in Figs.~1-5.
In each table, we list the number of SNIa in each subsample, the $95\%$ CL upper limit on the dipolar amplitude, and the $68\%$ CL constraints on the model parameters except the dipolar direction because of too poor constraint on it.
In each figure, we depict the marginalized probability distribution functions (PDFs) of all the six parameters for each subsample. Due to poor constraints on dipolar amplitude and direction, we conclude that there are no evidence for preferred directions in the universe.

\begin{table}
\scriptsize 
\caption{We list the $68\%$ CL constraints on the model parameters $\Omega_m$, $\alpha_{JLA}$ and $\beta_{JLA}$, the $95\%$ CL constraint on the model parameter $A_D$, and the number of SNIa in each subsample. }
\begin{tabular}{llllll}
  \hline
 $z$ bins &  $N_\textrm{SN}$ &  {$\Omega_m       $} &  {$\alpha_{JLA}   $}  & {$\beta_{JLA}    $} & {$10^3 A_D         $}  \\
  \hline
 $z$ $<$ 0.1 & 152 &  $< 0.282    $ &  $0.1460\pm 0.0120          $ & $2.970\pm 0.160            $ & $< 5.90       $ \\
 $z$ $\ge$ 0.1 & 588 & $0.282^{+0.051}_{-0.040}   $ & $0.1406\pm 0.0080  $ & $3.187\pm 0.098   $ & $< 1.72   $\\
 \hline
 $z$ $<$ 0.2	& 318 & $0.280^{+0.100}_{-0.120}      $ &$0.1403\pm 0.0085   $ & $3.010\pm 0.110 $ &$< 3.11 $ \\
 $z$ $\ge$ 0.2 & 422	&  $0.250^{+0.048}_{-0.054}   $ & $0.1460\pm 0.0110  $& $3.200\pm 0.120 $& $< 2.40 $ \\
\hline
$z$ $<$ 0.3	&	467 & $0.359\pm 0.083  $ & $0.1429\pm 0.0077   $ & $3.101\pm 0.099  $ & $< 2.18  $ \\
$z$ $\ge$ 0.3 &	273	 & $0.275^{+0.057}_{-0.068}   $ & $0.1380\pm 0.0140  $ & $3.110\pm 0.140 $ & $< 4.41  $\\
\hline
$z$ $<$ 0.4	&	525 & $0.315^{+0.064}_{-0.055}   $ & $0.1452\pm 0.0074   $ & $3.059\pm 0.094   $ & $< 1.89  $ \\
$z$ $\ge$ 0.4& 215 & $0.266^{+0.062}_{-0.088}   $& $0.1290\pm 0.0160     $& $3.240\pm 0.180   $& $< 4.04   $\\
\hline
$z$ $<$ 0.5	&	555 & $0.307^{+0.056}_{-0.042}   $ & $0.1438\pm 0.0070  $ & $3.124\pm 0.090    $ & $< 1.89  $ \\
$z$ $\ge$ 0.5 &	185& $0.264^{+0.068}_{-0.110} $& $0.1220\pm 0.0210  $& $3.030^{+0.190}_{-0.220}  $& $< 4.32 $\\
\hline
$z$ $<$ 0.6	&	595 	 & $0.312^{+0.049}_{-0.037}   $ & $0.1424\pm 0.0070   $& $3.151\pm 0.085   $ & $< 1.81    $ \\
$z$ $\ge$ 0.6 &145 & $0.222^{+0.060}_{-0.130} $ & $0.1370\pm 0.0240 $& $2.610^{+0.240}_{-0.300} $ & $< 4.64 $\\
\hline
$z$ $<$ 0.15                 & 226 & $< 0.194 $                                & $0.1360\pm 0.0100  $ & $3.020\pm 0.130$& $< 3.67 $\\
0.15 $\le$ $z$ $<$ 0.3 & 241 & $0.380^{+0.150}_{-0.220} $ & $0.1540\pm 0.0130 $ & $3.250\pm 0.170$ & $< 3.36         $ \\
$z$ $\ge$ 0.3               & 273    & $0.275^{+0.057}_{-0.068} $ & $0.1380\pm 0.0140 $& $3.110\pm 0.140 $& $< 4.41     $\\
\hline
$z <$ 0.15              & 226 	&  $< 0.194   $                      & $0.1360\pm 0.0100      $ & $3.020\pm 0.130              $ & $< 3.67 $ \\
0.15 $\le z <$ 0.25 & 175 & $0.320^{+0.086}_{-0.310}$ & $0.1570\pm 0.0140  $  & $3.250\pm 0.190             $ & $< 4.17     $ \\
0.25 $\le z <$ 0.5 &154  & $0.233^{+0.085}_{-0.150}    $ & $0.1490\pm 0.0150 $ & $3.230\pm 0.180             $ & $< 3.58    $ \\
 $z \ge$ 0.5         & 185  & $0.264^{+0.068}_{-0.110}    $& $0.1220\pm 0.0210 $& $3.030^{+0.190}_{-0.220} $& $< 4.32        $\\
\hline
$z <$ 0.1         &152 & $< 0.282                 $ & $0.1460\pm 0.0120 $ & $2.970\pm 0.160          $ & $< 5.90  $\\
0.1 $\le z <$ 0.2 &166 & $0.460^{+0.210}_{-0.290} $ & $0.1350\pm 0.0130 $ & $3.090^{+0.150}_{-0.170} $ & $< 4.86  $\\
0.2 $\le z <$ 0.3 &149 & $0.480^{+0.220}_{-0.400} $ & $0.1540\pm 0.0180 $ & $3.360\pm 0.220          $ & $< 3.98  $\\
0.3 $\le z <$ 0.6 &128 & $0.440^{+0.140}_{-0.190} $ & $0.1460\pm 0.0170 $ & $3.290^{+0.160}_{-0.180} $ & $< 5.63  $\\
$z \ge$ 0.6       &145 & $0.222^{+0.060}_{-0.130} $ & $0.1370\pm 0.0240 $ & $2.610^{+0.240}_{-0.300} $ & $< 4.64  $\\
\hline
\end{tabular}
\end{table}

\begin{figure}
  \centering
 \includegraphics[width=8 cm]{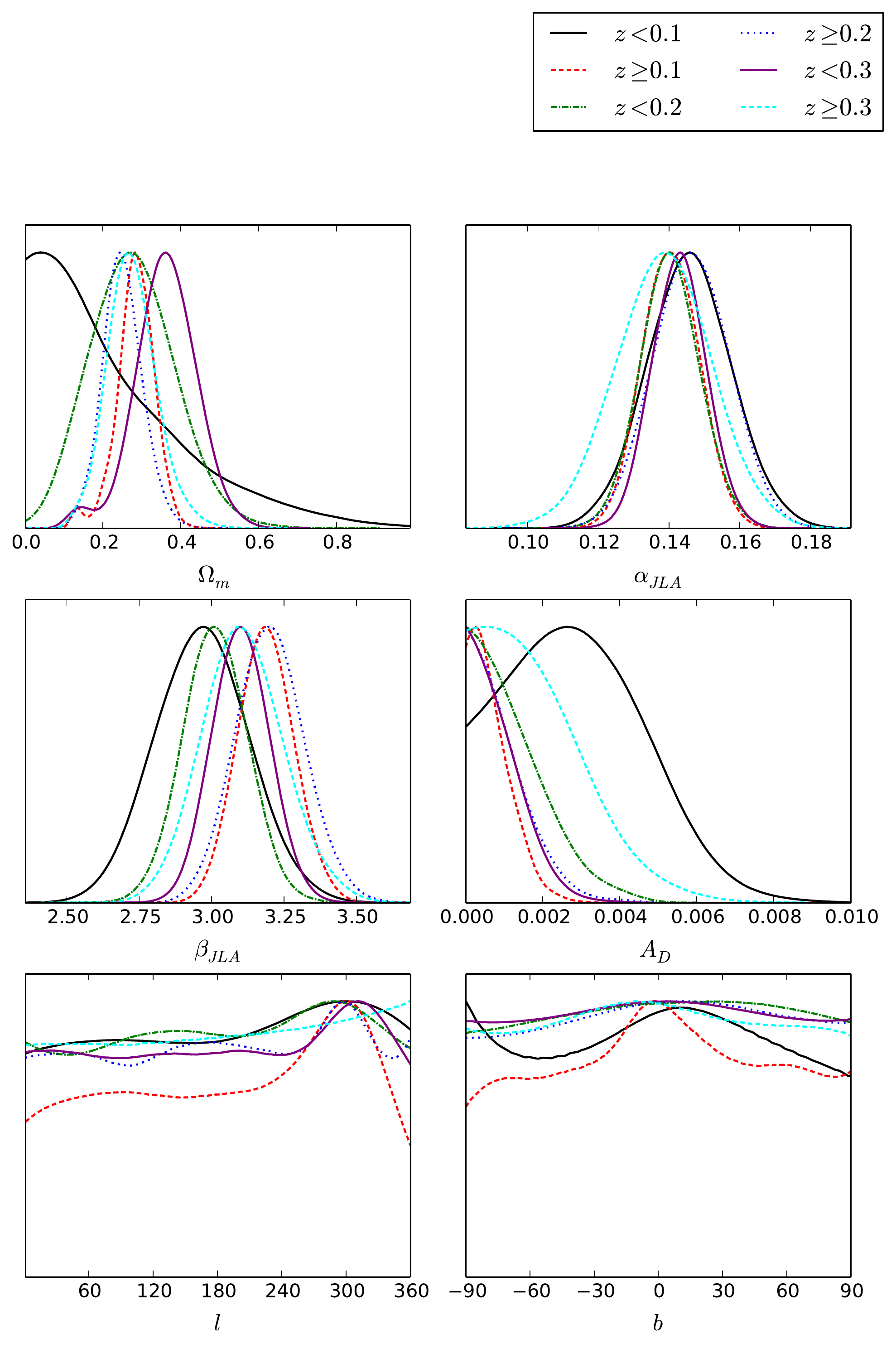}
 \caption{For the first, second and third tomography, we depict the marginalized probability distribution functions (PDFs) of all the six parameters for each subsample.}
\end{figure}

\begin{figure}
  \centering
 \includegraphics[width=8 cm]{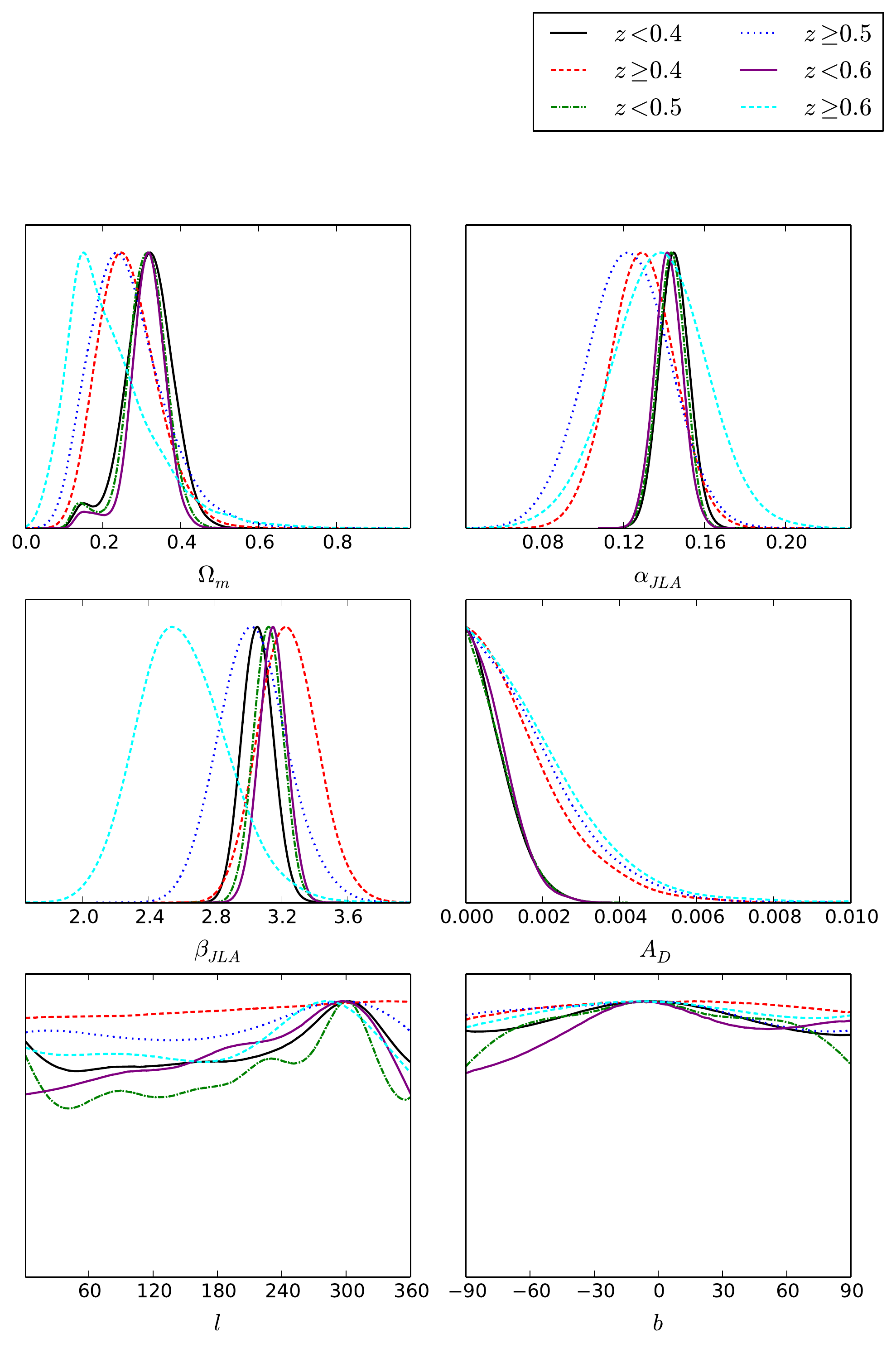}
 \caption{For the fourth, fifth and sixth tomography, we depict the marginalized probability distribution functions (PDFs) of all the six parameters for each subsample.}
 \label{fig:fig2}
\end{figure}

\begin{figure}
  \centering
 \includegraphics[width=8 cm]{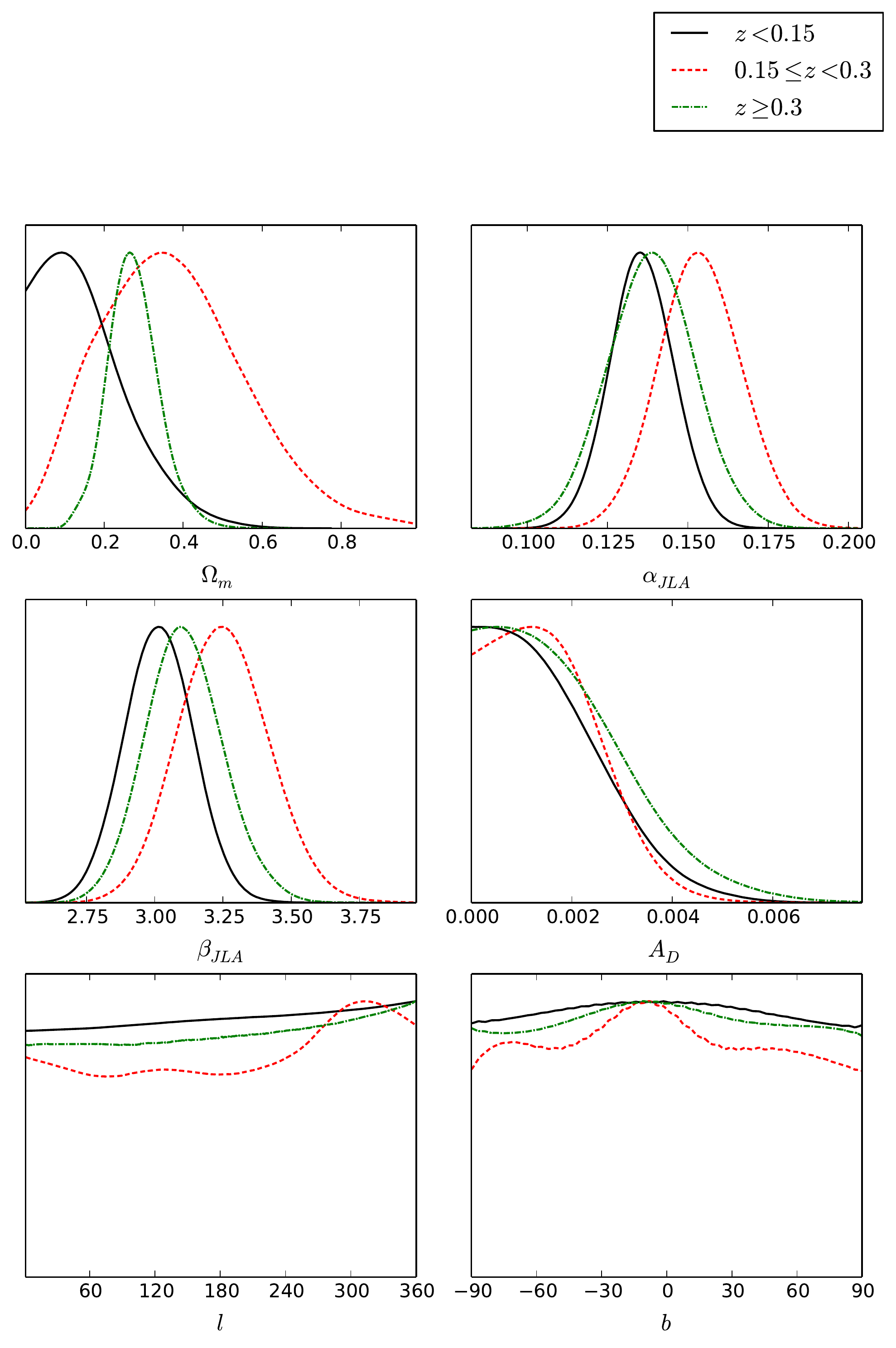}
 \caption{For the seventh tomography, we depict the marginalized probability distribution functions (PDFs) of all the six parameters for each subsample.}
\end{figure}

\begin{figure}
  \centering
 \includegraphics[width=8 cm]{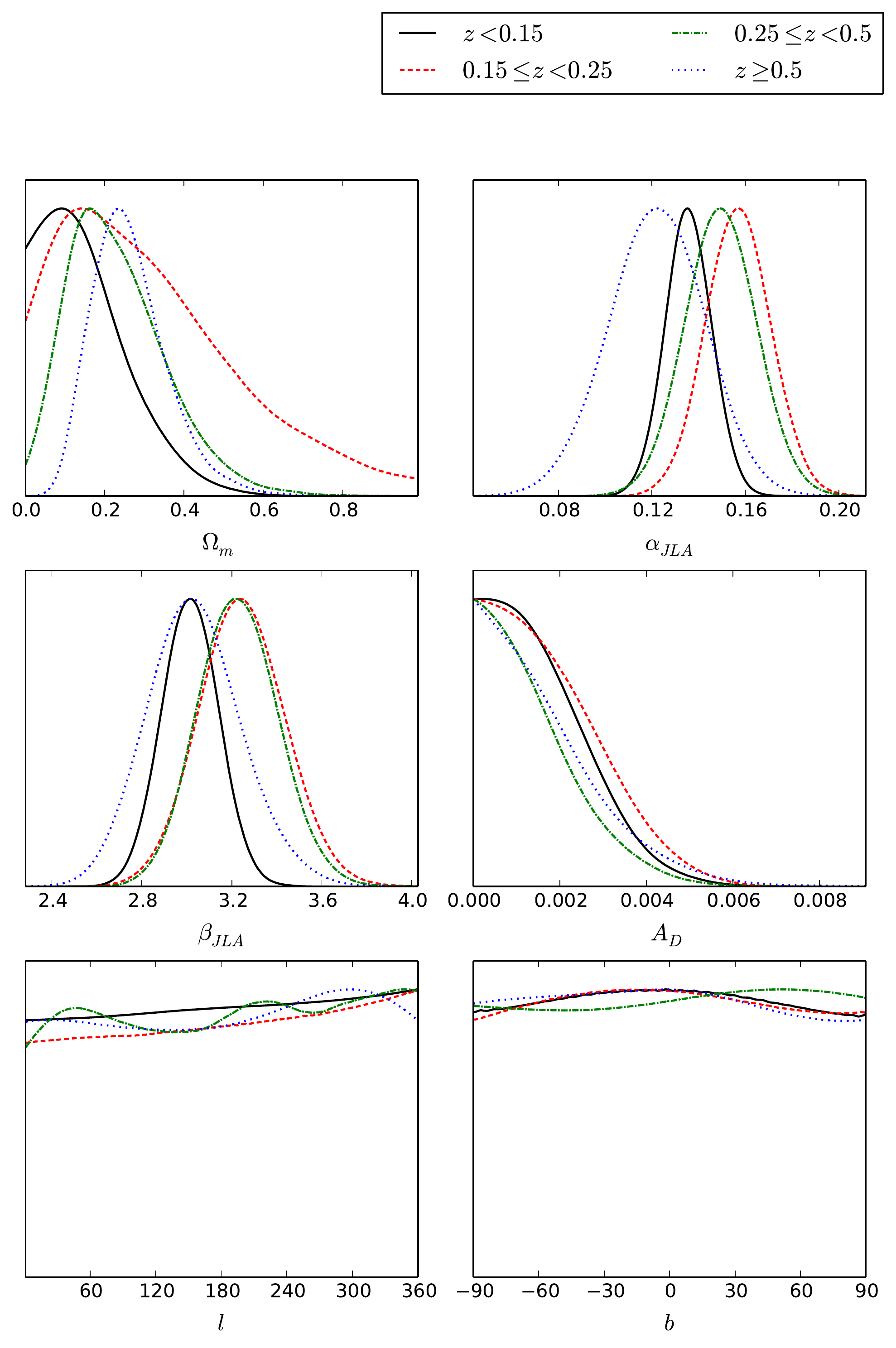}
 \caption{For the eighth tomography, we depict the marginalized probability distribution functions (PDFs) of all the six parameters for each subsample.}
\end{figure}


\begin{figure}
  \centering
 \includegraphics[width=8 cm]{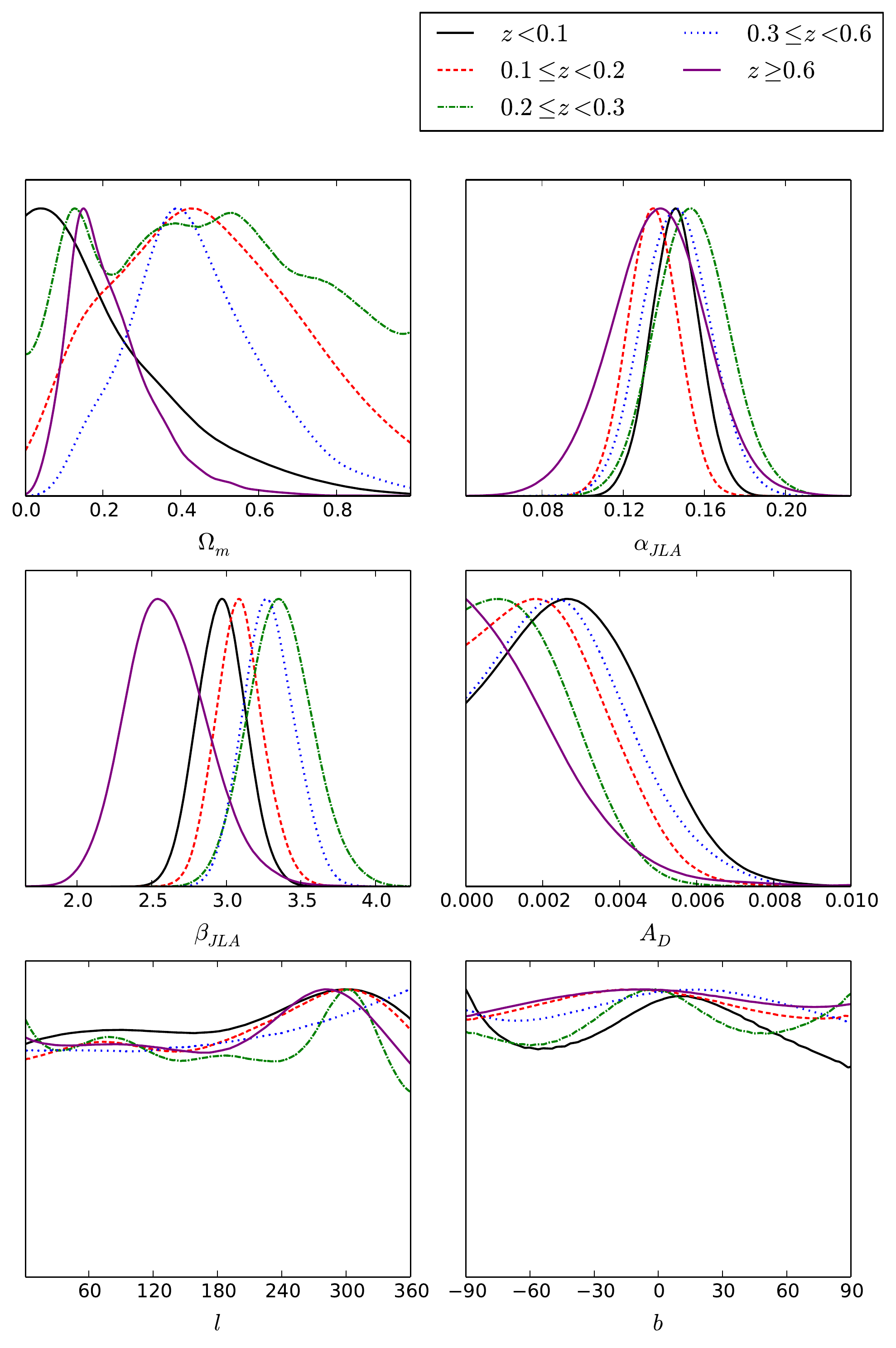}
 \caption{For the ninth tomography, we depict the marginalized probability distribution functions (PDFs) of all the six parameters for each subsample.}
 \label{fig:fig9}
\end{figure}

In the first tomography (see Tab. 1 and Fig. 1), due to a small number of SNIa, the lower redshift subsample ($z <$ 0.1) cannot give a strong constraint on $\Omega_\textrm{m}$. The values of  $\alpha$ given by two subsamples are consistent within 1$\sigma$ uncertainty. The parameter $\beta$ evolves slightly with redshift, and the value of $\beta$ given by the higher redshift subsample ($z \ge 0.1$) is larger than that given by the lower redshift one by about 1.4$\sigma$. The anisotropic amplitudes for lower and higher redshift subsamples are constrained as $A_D < 5.90\times10^{-3}$ and $A_D < 1.72\times10^{-3}$, respectively, at $95\%$ CL.

In the second tomography (see Tab. 1 and Fig. 1), both $\Omega_\textrm{m}$ and $\alpha$ constrained by two subsamples are consistent within 1$\sigma$ uncertainty. The parameter $\beta$ evolves slightly with redshift, and the value of $\beta$ given by the higher redshift subsample ($z \ge 0.2$) is larger than that given by the lower redshift one ($z <$ 0.2) by around 1.6$\sigma$. The anisotropic amplitudes for lower and higher redshift subsamples are constrained as $A_D < 3.11\times10^{-3}$ and $A_D < 2.40\times10^{-3}$, respectively, at $95\%$ CL.

In the third tomography (see Tab. 1 and Fig. 1), the value of $\Omega_\textrm{m}$ given by the higher redshift subsample ($z \ge 0.3$) is smaller than that given by the lower redshift one ($z <$ 0.3) by 1.0$\sigma$. Both $\alpha$ and $\beta$ constrained by two subsamples are consistent within 1$\sigma$ uncertainty. There is not significant evolution of model parameters with redshift. The anisotropic amplitudes for lower and higher redshift subsamples are constrained as $A_D < 2.18\times10^{-3}$ and $A_D < 4.41\times10^{-3}$, respectively, at $95\%$ CL.

In the fourth tomography (see Tab. 1 and Fig. 2), the values of $\Omega_\textrm{m}$ given by two subsamples are consistent within 1$\sigma$ uncertainty. The value of $\alpha$ given by the higher redshift subsample ($z \ge 0.4$) is smaller than that given by the lower redshift one ($z <$ 0.4) by $1.0\sigma$, while the value of $\beta$ given by the higher redshift one is larger than that given by the lower redshift one by $1.0\sigma$. The anisotropic amplitudes for lower and higher redshift subsamples are constrained as $A_D < 1.89\times10^{-3}$ and $A_D < 4.04\times10^{-3}$, respectively, at $95\%$ CL.

In the fifth tomography (see Tab. 1 and Fig. 2), both $\Omega_\textrm{m}$ and $\beta$ constrained by two subsamples are consistent within 1$\sigma$ uncertainty. The value of $\alpha$ given by the higher redshift subsample ($z \ge 0.5$) is smaller than that given by the lower redshift one ($z <$ 0.5) by $1.0\sigma$. The anisotropic amplitudes for lower and higher redshift subsamples are constrained as $A_D < 1.89\times10^{-3}$ and $A_D < 4.32\times10^{-3}$, respectively, at $95\%$ CL.

In the sixth tomography (see Tab. 1 and Fig. 2), the value of $\Omega_\textrm{m}$ given by the higher redshift subsample ($z \ge 0.6$) is smaller than that given by the lower redshift one ($z <$ 0.6) by 1.5$\sigma$. The values of $\alpha$ given by two subsamples are consistent within 1$\sigma$ uncertainty. The parameter $\beta$ evolves with redshift (see Fig.~\ref{fig:fig2}), and the value of $\beta$ given by the higher redshift subsample is smaller than that given by the lower redshift one by about $2.3\sigma$. The anisotropic amplitudes for lower and higher redshift subsamples are constrained as $A_D < 1.81\times10^{-3}$ and $A_D < 4.64\times10^{-3}$, respectively, at $95\%$ CL.

In the seventh tomography (see Tab. 1 and Fig. 3), the first subsample ($z <$ 0.15) cannot give a strong constraint on $\Omega_\textrm{m}$, although it includes 226 SNIa. The values of $\Omega_\textrm{m}$ given by the second subsample (0.15 $\le$ $z$ $<$ 0.3) and the third one ($z \ge 0.3$) are consistent within 1$\sigma$ uncertainty. The values of $\alpha$ given by the first subsample and the third one are consistent within 1$\sigma$ uncertainty. However, the second subsample gives a larger value of $\alpha$ than the first one by about $1.4\sigma$, and than the third one by $1.1\sigma$, respectively. The parameter $\beta$ evolves slightly with redshift, and the value of $\beta$ given by the second subsample is larger than that given by the first one by about $1.4\sigma$, and consistent with that given by the third one within 1$\sigma$ uncertainty. The anisotropic amplitudes are constrained by three subsamples as $A_D < 3.67\times10^{-3}$, $A_D < 3.36\times10^{-3}$ and $A_D < 4.41\times10^{-3}$, respectively, at $95\%$ CL.

In the eighth tomography (see Tab. 1 and Fig. 4), the first subsample ($z <$ 0.15) and the second one (0.15 $\le$ $z$ $<$ 0.25) cannot give a strong constraint on $\Omega_\textrm{m}$. The value of $\Omega_\textrm{m}$ given by the third subsample (0.25 $\le$ $z$ $<$ 0.5) and the fourth one ($z \ge 0.5$) are consistent within 1$\sigma$ uncertainty. Both of $\alpha$ and $\beta$ given by the lowest redshift subsample and highest redshift one show a slight trend of decrease compared to the two intermediate redshift subsamples. The value of $\alpha$ given by the second subsample is larger than that given by the first one by $1.5\sigma$. The value of $\alpha$ given by the third subsample is larger than that given by the fourth one by $1.3\sigma$. The values of $\alpha$ given by the second subsample and the third one are consistent within 1$\sigma$ uncertainty. The value of $\beta$ given by the second subsample is larger than that given by the first one by $1.2\sigma$. The value of $\beta$ given by the third subsample is larger than that given by the fourth one by around $1.1\sigma$. The values of $\beta$ given by the second subsample and the third one are consistent within 1$\sigma$ uncertainty. The anisotropic amplitudes are constrained by four subsamples as $A_D < 3.67\times10^{-3}$, $A_D < 4.17\times10^{-3}$, $A_D < 3.58\times10^{-3}$ and $A_D < 4.32\times10^{-3}$, respectively, at $95\%$ CL.

In the ninth tomography (see Tab. 1 and Fig. 5), the first subsample ($z <$ 0.1) and the third one (0.2 $\le$ $z$ $<$ 0.3) cannot give a strong constraint on $\Omega_\textrm{m}$. The value of $\Omega_\textrm{m}$ given by the second subsample (0.1 $\le$ $z$ $<$ 0.2) and the fourth one (0.3 $\le$ $z$ $<$ 0.6) are consistent within 1$\sigma$ uncertainty. The fifth subsample ($z \ge 0.6$) gives a smaller value of $\Omega_\textrm{m}$ than the fourth subsample (0.3 $\le$ $z$ $<$ 0.6) by around $1.2\sigma$. Except that the second subsample gives a smaller value than the third one by about $1.1\sigma$, the constraints on $\alpha$ given by other four subsamples are consistent with each other within 1$\sigma$ uncertainty. The constraints on $\beta$ given by the first two subsamples are consistent within 1$\sigma$ uncertainty. The value of $\beta$ given by the third subsample is larger than that given by the second one by $1.2\sigma$. The values of $\beta$ given by the third subsample and the fourth one are consistent within 1$\sigma$ uncertainty. The value of $\beta$ given by the fifth subsample is smaller than that given by the fourth one by around $2.8\sigma$. This may imply a significant evolution of $\beta$ with redshift (see Fig.~\ref{fig:fig9}). However, the anisotropic amplitudes are constrained by five subsamples to be $A_D < 5.90\times10^{-3}$, $A_D < 4.86\times10^{-3}$, $A_D < 3.98\times10^{-3}$, $A_D < 5.63\times10^{-3}$, and $A_D < 4.64\times10^{-3}$, respectively, at $95\%$ CL. These upper limits on $A_D$ are compatible among themselves.

\section{Conclusions}\label{sec:conclusions}

To test the cosmological principle, in this work, we fitted the dipolar modulation of distance modulus to the JLA compilation of supernovae using the redshift tomographic method. We did not find any significant deviations from the cosmological principle for all the nine cases of redshift tomography. The anisotropic amplitudes are thus stringently constrained to be less than a few thousandths at $95\%$ CL. These constraints are compatible with the one obtained in Ref.~\citep{Lin2016a} by non-use of the redshift tomographic method. Therefore, the universe is well consistent with the cosmological principle.

In the isotropic $\Lambda$CDM model, one has obtained $\Omega_m=0.295\pm0.034$, $\alpha=0.141\pm0.006$, and $\beta=3.101\pm0.075$ \citep{Betoule2014}, which are compatible with the results obtained in this work, except for a $2.1\sigma$ CL deviation from the constraint $\beta=2.610^{+0.240}_{-0.300}$ given by the subsample of $z\ge0.6$. Therefore, $\beta$ is redshift dependent, and has a trend of decrease at high redshift, while $\alpha$ remains constant in this work. Analyzing the JLA compilation, in fact, a previous work \citep{LiM2016} found a $\sim3.5\sigma$ CL evidence for the decrease of $\beta$ at high redshift in the isotropic $\Lambda$CDM model using the redshift tomographic method. However, it is still slightly different from our work, since the redshift tomographic method is not identical in both works. To be specific, all the model parameters are assumed to be piecewise constants in our redshift tomographic method, while the parameter $\Omega_m$ is assumed to be a global constant and others are piecewise constants in their method.

\section*{Acknowledgements}
This work has been funded by the National Natural Science Foundation of China under grant Nos.11375203, 11675182, 11690022, 11603005. SW is partially supported by a grant from the Research Grants Council of the Hong Kong Special Administrative Region, China (Project No. 14301214).



\bibliographystyle{mnras}
\bibliography{myreference} 








\bsp	
\label{lastpage}
\end{document}